\pdfoutput=1
\documentclass[a4paper,11pt]{article}
\usepackage[utf8]{inputenc}
\usepackage[left=2cm,right=2cm]{geometry}
\usepackage{tikz}
\usetikzlibrary{snakes}
\usepackage{slashed}
\usepackage{ulem}
\usepackage[font=small,labelfont=bf,width=1\textwidth]{caption}
\usepackage{subfigure}
\usepackage{comment}
\usepackage{yhmath}
\usepackage[all]{xy}
\usepackage{multirow}
\usepackage{float}
\usepackage{mathtools}
\usepackage{xfrac}
\usepackage{wrapfig}
\usepackage{dsfont}
\usepackage{makecell}
\usepackage{color}
\usepackage{bbding}
\usepackage{pifont}
\usepackage{wasysym}
\usepackage{amssymb}
\usepackage{subcaption}
\usepackage{cite}
\usepackage[symbol]{footmisc}
\usepackage[colorlinks=true,linkcolor=black,citecolor=blue,urlcolor=blue]{hyperref}
\numberwithin{equation}{section}

\DeclareMathOperator{\Tr}{Tr}

\title{
From 2D Yang-Mills to Calogero-Sutherland via a colored particle
}
\date{}

\begin{document}

\maketitle
\vspace{-1.75cm}
\begin{center}
{\large Marcia Tenser$^a$\footnote[2]{marciatenser@gmail.com} and Amilcar R. Queiroz$^b$\footnote[3]{amilcarq@gmail.com}}
\\
~
\\
\textit{\small $^a$International Institute of Physics, Universidade Federal do Rio Grande do Norte,\\ 59078-970, Natal, RN, Brazil\\
\vspace{0.2cm}
$^b$Departamento de F\'{\i}sica, Universidade Federal de Campina Grande,\\
Caixa Postal 10071, 58429-900, Campina Grande, Para\'{\i}ba, Brazil
}

\vspace{2cm}

\end{center}

\vspace{-2cm}

\begin{abstract}
     We study Yang-Mills theory coupled to a particle on a cylinder, where gauge invariance and compactness reduce the dynamics to a finite dimensional quantum system. In the Abelian case, this yields a model equivalent to the Landau problem on a torus, with a degenerate ground state structure. We generalize this construction to non-Abelian gauge groups and show that, for $SU(N)$, the system reduces to a one dimensional quantum many body problem with a singular Calogero-Sutherland-type interaction.
\end{abstract}

\tableofcontents


\maketitle

\let\clearpage\relax
\section{Introduction}

Among various approaches to studying gauge theories, lower-dimensional models on compact spaces are particularly valuable. Two-dimensional Yang-Mills theory on a cylinder, for example, carries no local propagating degrees of freedom. Its dynamics is completely reduced to a finite set of global, quantum-mechanical variables associated with holonomies around the compact direction. This reduction maps a quantum field theory problem onto one of quantum mechanics, offering a clear setting to study confinement, topological sectors, and the role of large gauge transformations \cite{Rajeev:1988zb, Hetrick:1989vm, Witten:1991we, Cordes:1994fc}.

In this context, further richness may be added to the system by coupling dynamical matter to Yang-Mills fields. This eventually opens new avenues of investigation, including the study of entanglement entropy in the presence of gauge symmetry.

In a previous work~\cite{Ares:2019oil}, we considered a non-relativistic charged particle coupled to Maxwell theory on a cylinder. A careful treatment of gauge symmetry and the compactness of the spatial direction reduced the system to a quantum-mechanical model with two degrees of freedom, equivalent to the Landau problem on a torus \cite{thouless1982,Carvalhoetal2014}. This mapping allowed for an exact computation of the entanglement entropy between the particle and the gauge field. The entropy was found to contain a contribution proportional to the axial anomaly \cite{Manton:1985jm, Esteve:1986db, Aguadoetal2001} of the 2D theory, highlighting an interplay between quantum information measures and topological aspects of gauge theories.

Naturally, the results found in~\cite{Ares:2019oil} motivate the extension of the entanglement analysis to non-Abelian groups, where the anomaly structure is richer and the Hilbert space exhibits a more intricate factorization. Such a model offers a tractable non-Abelian setting in which to explore how colored particle degrees of freedom entangle with the global holonomies of the gauge field. Our ultimate goal is therefore to derive exact expressions for the entanglement entropy, thereby generalizing the Abelian results of~\cite{Ares:2019oil} and shedding light on the quantum information content of non-Abelian gauge theories.

In this paper, we take a first step toward this goal by deriving the exact gauge-reduced Hamiltonian of the particle coupled with 2D Yang-Mills. Our approach follows the Hamiltonian framework developed by Langmann and Semenoff \cite{Langmann:1992np, Langmann:1993pb, Hetrick:1993pa}, in which gauge redundancies are eliminated by explicitly solving the Gauss law constraint prior to quantization. While the authors consider Yang-Mills theory coupled to massless Dirac fermions, here we couple the gauge field to a non-relativistic point particle carrying a non-Abelian color charge. Our modification leads to a Hamiltonian that retains the essential features of the Langmann-Semenoff construction but replaces the fermionic degrees of freedom with the particle's position and isospin.\footnote{The approach of gauge-reducing and then quantizing is complementary to the strategy employed by Rajeev and collaborators \cite{Rajeev:1988zb, Gupta:1993gg}, who first quantize and subsequently impose gauge invariance at the quantum level. Both methods have their merits; the former offers a more direct route to the physical Hilbert space and reveals the underlying integrable structure, while the latter keeps gauge invariance manifest throughout.}

We focus on the derivation of the exact Hamiltonian governing the system and find that, despite its conceptual simplicity, the elimination of gauge redundancies in the presence of a dynamical non-Abelian charge is technically nontrivial. In carrying out this reduction, we uncover a direct connection between our system and the Calogero-Sutherland model \cite{Calogero:1970nt, Sutherland:1971kq, Sutherland:1972, Olshanetsky:1983}, a result that we believe merits attention in its own right.

Our main result is the derivation of the gauge-reduced Hamiltonian (3.26), which governs the dynamics of a color-charged particle coupled to $SU(N)$ Yang-Mills theory on a cylinder. In this reduced description, the system behaves as a one-dimensional gas of $N$ interacting particles: the original dynamical particle plus $N-1$ effective particles associated with the gauge holonomies. The interaction between these particles is of the Calogero-Sutherland type \cite{Calogero:1970nt, Sutherland:1971kq, Sutherland:1972, Olshanetsky:1983}, with a potential determined by the root system $A_{N-1}$ of the gauge group, supplemented by an additional harmonic term arising from the minimal coupling between the particle's position and the gauge field. This mapping provides a transparent example of how non-Abelian gauge dynamics can be recast in the language of quantum integrable systems, a correspondence that has been explored in related contexts \cite{Minahan:1993,Minahan:1993mv, Gorsky:1994}.

\vspace{0.5cm}

The paper is organized as follows. In Section~\ref{sec:EOM}, we present the physical setup of Yang-Mills theory coupled to a non-relativistic particle on a cylinder and derive the classical equations of motion. Section~\ref{sec:Hamiltonian} is devoted to the Hamiltonian formulation. We solve the Gauss law constraint by rotating to the Cartan basis, a procedure that systematically eliminates gauge redundancies and yields the gauge-reduced Hamiltonian. Section~\ref{sec:calogero} makes explicit the relation between this dynamics and that of a Calogero-Sutherland model. The quantization of this system is treated in Section~\ref{sec:Quantization}, where we discuss the structure of the allowed wavefunctions on the space of gauge orbits and illustrate the results for the $SU(2)$ and $SU(3)$ gauge groups. We conclude in Section~\ref{sec:discussion} with a summary and an outlook on future directions, including the prospects for studying entanglement between the matter and gauge sectors in this non-Abelian setting. Appendix~\ref{app:grouptheory} collects our group-theoretic conventions. The technical core of the paper, namely the detailed derivations of the results presented in Section~\ref{sec:Hamiltonian}, in particular the solution of the Gauss law constraint, is given in Appendix~\ref{app:gauss}. This material is included as an appendix to streamline the presentation of the main text. Finally, Appendix~\ref{app:weyl} details the action of the Weyl group and demonstrates the invariance of the Hamiltonian.


\section{Yang-Mills coupled to a non-relativistic particle}
\label{sec:EOM}

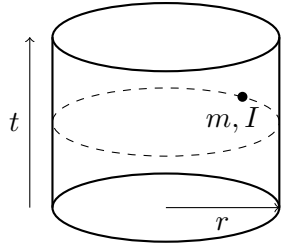
\begin{figure}[H]
    \centering
    \begin{tikzpicture}[scale=0.75]
      \def\R{2}    
      \def\t{3}    
      
      \draw[thick] (0,0) ellipse (\R cm and 0.6cm);
      \draw[thick] (-\R,0) -- (-\R,\t);
      \draw[thick] (\R,0) -- (\R,\t);
      \draw[thick] (0,\t) ellipse (\R cm and 0.6cm);

      \draw[dashed] (0,\t+1.4cm) ellipse (\R cm and 0.6cm);

      \draw[->] (0,0) -- (\R,0) node[midway,below] {$r$};

      \coordinate (A) at (1.35,1.95); 
      \fill (A) circle (2.5pt) node[below] {$m,I\,\,\,\,$};

      \draw[->] (-2.4,0) -- (-2.4,\t) node[midway,left] {$t$};
    \end{tikzpicture}
    \caption{The setup: a cylindrical manifold $\mathcal{M}=\mathbb{R}\times S^1$ of radius $r$; a point-particle of mass $m$ and isospin $I$. Time evolution is in the vertical direction.}
    \label{fig:cylinder&A}
\end{figure}

The model is Yang-Mills theory coupled to a non-relativistic charged point particle. We consider a cylindrical manifold $\mathcal{M}=\mathbb{R}\times S^1$ equipped with a Minkowski metric $\eta^{\mu\nu}=\rm{diag}(-1,1)$, with coordinates $\mathbf{x}=(x^0,x^1)\equiv (t,x)$ where $t\in\mathbb{R}$ and $x\in [0,2\pi r)$. The charged particle has mass $m$, its color charge (also called isospin or isotopic spin) is $I$ and its trajectory is described by ${\bf z}(\tau)=(z^0(\tau),z^1(\tau))\equiv (z^0(\tau),z(\tau))$ where $\tau$ parametrizes its wordline. The Yang-Mills coupling is denoted as $g_{\rm YM}$. The system may be depicted as shown in Figure \ref{fig:cylinder&A}.

The internal symmetry group $G$ is a compact, connected Lie group. We focus on the particular case of $G=SU(N)$. Notation and conventions regarding the gauge group structure are outlined in Appendix \ref{app:grouptheory}. 

The dynamics of the charged point-particle is described by a set of equations originally derived by Wong \cite{Wong:1970fu},
\begin{align}
        D_\mu F^{\mu\nu}&=g_{\rm YM}^2 \int d\tau \, \dot{z}^\nu(\tau) I(\tau) \delta^{(2)}({\bf x}-{\bf z}(\tau))\,,\label{eq:Wong1}\\
        m\ddot z^\mu &= 2{\rm Tr}(I F^{\mu\nu})\dot z_\nu\,, \label{eq:Wong2}
\end{align}
where the over dot stands for differentiation with respect to $\tau$. 

The quantity appearing on the right-hand side of \eqref{eq:Wong1} is the non-Abelian vector current
\begin{equation}\label{eq:Jvector}
    J^\mu({\bf x}) \equiv  \int d\tau \, \dot{z}^\mu(\tau) I(\tau) \delta^{(2)}({\bf x}-{\bf z}(\tau))\,.
\end{equation}
It is a Lie-algebra valued quantity since $I(\tau)= I^a(\tau)T^a$ is in the adjoint representation of the gauge group. From \eqref{eq:Wong1} and the identity $[D_\mu,D_\nu]F^{\mu\nu}=0$, one can show that the current is covariantly conserved,
\begin{equation}\label{eq:DJ}
    D_\mu J^\mu = 0\,.
\end{equation}

Using the definition of the vector current, we rewrite  \eqref{eq:DJ} as 
\begin{equation}
    \int d\tau\, \left(\frac{d(I(\tau))}{d\tau}-i[\dot{z}^\mu A_\mu,I(\tau)]\right)\delta(x-z(\tau))\delta(t-z^0(\tau)) =0.
\end{equation}
so that the isotopic spin satisfies
\begin{equation}\label{eq:DtauI}
    D_\tau I = \frac{dI}{d\tau}-i[\dot z^\mu A_\mu,I]=0\,.
\end{equation}
That is, it precesses in the internal space.

In two-dimensions, the Yang-Mills term contains only the electric-like field, $E\equiv F_{01}= \partial_0A_1-\partial_1 A_0-i[A_0,A_1]$. Aligning $\tau$ with the time coordinate $t$, the vector current \eqref{eq:Jvector} yields
\begin{align}\label{eq:rhoandj}
    \begin{split}
        \rho(t,x)&\equiv J^0(t,x)= I(t)\delta(x-z(t)),\\
        j(t,x)&\equiv J^1(t,x)= \dot{x}\, \rho(t,x)\,.
    \end{split}
\end{align}
The Wong's equations \eqref{eq:Wong1} and \eqref{eq:Wong2} become
\begin{align}\label{eq:EoM}
    \begin{split}
        \frac{1}{g_{\rm YM}^2}\bigg(\partial_t E - i[A_0,E] \bigg)&= -j, \\
        \frac{1}{g_{\rm YM}^2}\bigg( \partial_x E - i[A_1,E]\bigg) &= \rho ,\\
        m\ddot z &= 2{\rm Tr}\left(I E\right),
    \end{split}
\end{align}
where the over dots now stand for differentiation with respect to $t$.
\section{The Hamiltonian description}
\label{sec:Hamiltonian}

The Hamiltonian $H$ describing the system (\ref{eq:EoM}) contains the Yang-Mills contribution $H_{\rm YM}$ and the particle's contribution $H_{\text p}$, $H = H_{\rm YM} + H_{\rm p}$. 
The former takes its usual form in two spacetime dimensions \cite{Gupta:1993gg},
\begin{align}
\begin{split}
     H_{\rm YM} &= \frac{1}{2g_{\rm YM}^2}\int_{S^1} dx \, {\rm Tr}\bigg[ E^2 -2 A_0 \big(\partial_x E - i[A_1,E]\big)\bigg]\\
     &=\frac{1}{2}\int_{S^1} dx\, {\rm Tr}\bigg[ g_{\rm YM}^2\Pi^2 -2 A_0 \big(\partial_x \Pi - i[A_1,\Pi]\big)\bigg],
\end{split}
\end{align}
where $\Pi\equiv\frac{E}{g_{\rm YM}^2}$ is the canonical conjugate pair of $A_1$. 
The latter, on the other hand, is given by \cite{Balachandran:1977ub}
\begin{align}\label{eq:Hp}
    \begin{split}
        H_{\text p} &= \frac{1}{2m}\big(p_z- {\rm Tr}(A_1(z)  I)\big)^{2} + {\rm Tr}\big(A_0(z)  I\big),
    \end{split}
\end{align}
where $p_z=m\dot z+{\rm Tr}(A_1I)$. Thus, the full Hamiltonian may be written as
\begin{equation}\label{eq:fullH}
    H =  \frac{1}{2m}\big(p_z- {\rm Tr}(A_1(z) I)\big)^{2} + \frac{1}{2}\int_{S^1} dx\, {\rm Tr}\bigg[g_{YM}^2\Pi^2 -2 A_0 \big(\partial_x \Pi - i[A_1,\Pi] - \rho(x,t)\big)\bigg] \,,
\end{equation}
where we used that $A_0(z)I = \int dx \,A_0(x)\rho(x,t) $ to include the last term of \eqref{eq:Hp} inside of the integration. It is straightforward to derive the equations of motion \eqref{eq:EoM} from the Hamiltonian \eqref{eq:fullH}.

\subsection{Coulomb gauge}

The time component of the gauge field, $A_0$, works as a Lagrange multiplier and the last term in \eqref{eq:fullH} is a constraint. This means that we have
\begin{equation}\label{eq:Gauss}
    G \equiv \partial_x \Pi - i [A_1,\Pi] -\rho\simeq 0\,,
\end{equation}
which is simply the (1+1)d version of the Gauss law. Here, $\simeq$ denotes a weak equality, meaning that the constraint holds on the subspace of physical configurations. We solve such a constraint following \cite{Langmann:1992np} before quantizing the system. 

We introduce the Wilson line $S(x)$, which is the solution to
\begin{align}\label{eq:WLequations}
    \begin{split}
        (\partial_x -i A_1) S&=0\,,\\
        S(x=0)&=1\,.
    \end{split}
\end{align}
In terms of $S(x)$, tilded variables may be defined by conjugate action; that is, $\tilde \Pi(x)=S(x)^{-1}\Pi(x)S(x)$, and similarly for other quantities. In terms of these, \eqref{eq:Gauss} becomes
\begin{equation}
    \partial_x\tilde{\Pi}=\tilde\rho\,,
\end{equation}
which can be readily solved by
\begin{equation}
     \tilde \Pi(x) - \tilde{\Pi}(0) = \int_0^x dx' \tilde \rho(x') \equiv R(x)\,.
\end{equation}
Integrating this expression over $x$ between $0$ and $2\pi r$ and reinserting $\tilde{\Pi}(0)=\tilde{\Pi}(x)-R(x)$ back into the result, gives
\begin{equation}\label{eq:tildeE}
    \tilde{\Pi}(x)= P +\bar{R}(x),
\end{equation}
where
\begin{equation}
    P\equiv \frac{1}{2\pi r} \int_0^{2\pi r} dx\, \tilde{\Pi}(x)\,, \qquad \bar{R}(x)\equiv R(x)-\frac{1}{2\pi r} \int_0^{2\pi r} dx' R(x')\,.
\end{equation}
Now, recalling that $\tilde \Pi(x)=S(x)^{-1}\Pi(x)S(x)$ and $\Pi(2\pi r)=\Pi(0)$, it follows that
\begin{align}
\begin{split}
    \tilde \Pi(2\pi r)=S(2\pi r)^{-1}\Pi(2\pi r)S(2\pi r)=S(2\pi r)^{-1}\Pi(0)S(2\pi r)= P+ \bar R(2\pi r),
\end{split}
\end{align}
where, to write the last equality, we used \eqref{eq:tildeE} evaluated at $x=2\pi r$. Finally, noticing that $\Pi(0)=\tilde \Pi(0)=P+\bar R(0)$ and defining $S(2\pi r)\equiv q$, we obtain
\begin{equation}\label{eq:equivarianceconstraint}
    P+ \bar R(2\pi r) = q^{-1}\left(P+\bar R(0) \right) q\,.
\end{equation}
Note that $q$ is nothing but the Wilson loop around the circle. In particular, it can be shown that $P$ and $q$ satisfy the commutation relations
\begin{equation}\label{eq:pqcommutator}
    [P^a,q]=-qT^a\,, \qquad [P^a,q^{-1}]=T^a q^{-1}\,.
\end{equation}

In summary, we have traded the Gauss law constraint for a conjugate variable $\tilde{\Pi}(x)$, see \eqref{eq:tildeE}, where $P$ and $\bar{R}(x)$ satisfy \eqref{eq:equivarianceconstraint}. The Hamiltonian of the system becomes
\begin{equation}\label{eq:Htilded}
    H = \frac{1}{2m}\big(p_z-{\rm Tr}(A_1(z) I)\big)^2 + \frac{g_{\rm YM}^2}{2} \int_{S^1} dx \,{\rm Tr}\big(\tilde{\Pi}^2(x)\big) \,.   
\end{equation}
We will next solve \eqref{eq:equivarianceconstraint} by rotating to the Cartan subalgebra.

\subsection{Rotating to the Cartan basis}
\label{subsec:cartanbasis}

The Hamiltonian \eqref{eq:Htilded} and the relation \eqref{eq:equivarianceconstraint} are invariant under $x$-independent gauge transformations $U(t)=\exp(-i\Lambda(t))$. Under these, the Wilson loop $q$ transforms as
\begin{equation}
    q \rightarrow \tilde{q} = Uq\, U^{\dagger}\,.
\end{equation}
Moreover, every element $q$ of the gauge group $G$ may be represented as $q=h_q^{-1} \mathfrak{q} h_q$ with $\mathfrak{q}$ in the Cartan subgroup of $G$, and $h_q\in G$ defined up to right multiplication by elements of the Cartan subgroup. These facts together motivates the choice $U=h_q$, which is such that we rotate the system to the Cartan basis. Following the notation outlined in Appendix \ref{app:grouptheory}, this means that we may write $\mathfrak{q}$ in terms of Cartan generators $H^i$, $i=1,\dots,N-1$, as\footnote{Throughout the text, unless explicitly stated, repeated indices are assumed to be summed over.}
\begin{equation}\label{eq:qCartan}
    \mathfrak{q}= e^{2\pi i Y^i(t) H_i}\,.
\end{equation}
Recalling that $q$ is the Wilson loop variable, \eqref{eq:qCartan} implies that the Wilson line now reads $S(x)=e^{i Y^i H_i x/r}$ and the gauge field reads 
\begin{equation}\label{eq:Acartan}
    A_1=\frac{Y^i(t)H_i}{r}\,.
\end{equation}
This is essentially the Coulomb gauge.

The rotation to the Cartan basis has non-trivial consequences. For simplicity, here we present the main results, leaving the computational details to the interested reader in Appendix \ref{app:gauss}.

First, using \eqref{eq:Acartan} and decomposing the isotopic spin $I(t)$ in the Cartan-Weyl basis, 
\begin{equation}
    I(t)= I_0^{i}(t)H_i+I_+^{\hat\jmath}(t)E_{\hat\jmath}^- +I_-^{\hat\jmath}(t)E_{\hat\jmath}^+\,,
\end{equation}
we readily solve its equation of motion \eqref{eq:DtauI}. Cartan components are constant,
\begin{equation}
        I_0^{i}(t)= Q^{i}_0\,,
        \label{eq:Isol1}
\end{equation}
while root components (labeled by $\hat{\imath},\,\,\hat{\imath}=1,\dots,N(N-1)/2$) oscillate over time,
\begin{equation}
       I_\pm^{\hat\imath}(t)=Q_\pm^{\hat\imath} \exp\left({\mp \frac{i}{r} \int_0^t a_{i\hat\imath}Y^i(t')\dot z\,dt'}\right)\,.\label{eq:Isol2}
\end{equation}
The elements $a_{i\hat{\imath}}$ are the components of the root matrix; see Appendix~\ref{app:grouptheory}. In \eqref{eq:Isol2} there is no implicit sum over $\hat\imath$. 

In the same spirit, the decomposition of $\bar{R}$ and $P$ in the Cartan-Weyl basis, together with \eqref{eq:equivarianceconstraint}, allows us to relate their components and ultimately find simple expressions for $\tilde{\Pi}$, which appears in the Hamiltonian \eqref{eq:Htilded} (recall that $\tilde{\Pi}=P+\bar{R}(x)$). In particular, $P_0$ is left unconstrained, while
\begin{equation}
    \bar R_0^j(x)=-i\sum_{n=-\infty}^{\infty} \varrho_0^{j}(n)\frac{e^{inx/r}}{2\pi n}(1-\delta_{n,0})\,,
\end{equation}
and
\begin{equation}\label{eq:PR_hatj}
        P_\pm^{\hat\jmath} + \bar{R}_\pm^{\hat\jmath}(x) = -i \sum_n \varrho_\pm^{\hat\jmath}(n)\frac{ e^{i(n\mp  Y^ia_{i{\hat\jmath}})x/r}}{2\pi(n\mp Y^ia_ {i{\hat\jmath}})}\,.
\end{equation}
Here $\varrho(n)$ is the Fourier transform of the charge distribution $\rho(x)$,
\begin{equation}
    \varrho(n) =\int_0^{2\pi r} dx \,\rho(x) e^{-inx/r}\,,
\end{equation}
which was expanded in the Cartan-Weyl basis. There is no sum over $\hat\jmath$ in the right hand side of \eqref{eq:PR_hatj}.

The charge distribution at hand, $\rho(t,x)= I \delta(x-z)$, then gives
\begin{align}\label{eq:trEtilde3}
    \begin{split}
        \int_0^{2\pi r}dx\, {\rm Tr}(\tilde{\Pi}^2)
        &= 2\pi r P_0^ib_{ij}P_0^j + \frac{\pi r}{2} \left(\frac{1}{3} Q_0^{i}b_{ij}Q_0^{j} + \frac{Q_+^{{\hat\imath}}Q_-^{{\hat\imath}}+Q_-^{{\hat\imath}}Q_+^{{\hat\imath}}}{\sin^2(\pi Y^i a_{i{\hat\imath}})}\right)\,,
    \end{split}
\end{align}
where $b_{ij}$ are the elements of the Cartan matrix; again, definitions are in Appendix~\ref{app:grouptheory}. Moreover, the term ${\rm Tr}(A_1(z) I)$ appearing in the Hamiltonian \eqref{eq:Htilded} becomes
\begin{equation}
\begin{split}
    {\rm Tr}\big(A_1(z) I\big)=\frac{1}{r}Y^i(t) b_{ij}Q_0^j  \,.
\end{split}
\end{equation}
Thus the full Hamiltonian of the system takes the form
\begin{equation}
H = 
\frac{1}{2m}\left(p_z-\frac{1}{r}Y^iQ_{0i}\right)^2
+ g_{\rm YM}^2 \pi r\, P_0^iP_{0i}
+ \frac{g_{\rm YM}^2\pi r}{4}
\bigg(
\frac{Q_0^{i}Q_{0i}}{3}
+ \frac{Q_+^{{\hat\imath}}Q_-^{{\hat\imath}}+Q_-^{{\hat\imath}}Q_+^{{\hat\imath}}}
{\sin^2(\pi Y^i a_{i{\hat\imath}})}
\bigg)
\end{equation}
where $Q_{0i}\equiv \displaystyle\sum_{j=1}^{N-1} b_{ij}Q_0^{j}$ and similarly for $P_{0i}$. 

Defining
\begin{equation}\label{eq:KjandmYM}
     m_{\rm YM}\equiv \frac{1}{g_{\rm YM}^2\pi r}\,, \qquad  K^{\hat\imath}\equiv \frac{Q_+^{{\hat\imath}}Q_-^{{\hat\imath}}+Q_-^{{\hat\imath}}Q_+^{{\hat\imath}}}{2}\,, \qquad V(\mathbf{Y}) = \frac{K^{\hat{\imath}}}{\sin^2(\pi Y^i a_{i\hat{\imath}})}
\end{equation}
and neglecting the constant term proportional to $Q_0^{i}Q_{0i}$, the final expression for the Hamiltonian is
\begin{equation}\label{eq:finalH}
    H=\frac{1}{2m}\left(p_z-\frac{1}{r}Y^i Q_{0i}\right)^2+\frac{1}{m_{\rm YM}}P_{0}^i P_{0i}+\frac{1}{2m_{\rm YM}}V(\mathbf{Y})\,.
\end{equation}
We stress that indices $i$ run from 1 to $N-1$ and ${\hat\jmath}$ from 1 to $N(N-1)/2$. The resulting dynamics correspond to that of a gas of $(N-1)+(1)$ particles in their respective domains. Their coordinates are $\mathbf{Y}=\{Y^1,\dots,Y^{N-1}\}$ and $z$. The system was reduced to a one dimensional many-body problem.

The first two elements in the Hamiltonian \eqref{eq:finalH} carry a Cartan-matrix dependence, namely via $Q_{0}^i$ and $P_{0}^i$. These are related to the Abelian substructure of the group. Naturally, such elements were already present in our previous studies~\cite{Ares:2019oil}, where a charged particle was coupled to a Maxwell theory on a cylinder, the difference being that here they appear with $(N-1)$-multiplicity. 

The potential term $V(\mathbf{Y})$, on the other hand, has no Abelian analog and is ``root system" dependent. Below we show that this interaction term is of the Calogero-Sutherland type \cite{Calogero:1970nt,Sutherland:1971kq}.


\subsection{The Calogero-Sutherland interaction}
\label{sec:calogero}

Although \eqref{eq:finalH} captures the essential interaction, its form obscures the characteristic pairwise structure of the Calogero-Sutherland model. To make it explicit, we change from the Cartan subalgebra basis $\{\mathbf{Y}\}$ to the basis of holonomy eigenvalues $\{\mathbf{y}\}$. These are related by the linear transformation
\begin{equation}\label{eq:Y_as_sum_of_y}
Y^i = \sum_{m=1}^{i} y^m\,, \qquad i = 1, \dots, N-1\,,
\end{equation}
together with the traceless condition $\sum_{m=1}^N y^m = 0$.

In the eigenvalue basis, the arguments of the sine functions in the potential simplify. For each positive root, the combination $Y^i a_{i\hat{\imath}}$ reduces to the difference between two eigenvalues, namely, $y^m-y^n$. The corresponding $K^{\hat\imath}$ in the potential is labeled by the corresponding root pair, that is, $K_{mn}$.\footnote{See Appendix~\ref{app:weyl} and, in particular, the discussion around equation \eqref{eq:permutation_K} for the relevant group-theoretical details.} Consequently, the potential assumes the canonical form of a pairwise interaction,
\begin{equation}\label{eq:CS-potential}
V(\mathbf{y}) = \sum_{1 \le m < n \le N} \frac{K_{mn}}{\sin^2\!\big(\pi(y^m - y^n)\big)}\,. 
\end{equation}
This is precisely the interaction potential of the Calogero-Sutherland model of type $A_{N-1}$, which describes $N$ particles repelling each other through an inverse sine-square potential \cite{Calogero:1970nt, Sutherland:1971kq, Olshanetsky:1983, Minahan:1993}. In the Cartan classification, $A_{N-1}$ denotes the root system of the special unitary group $SU(N)$, confirming that the non-Abelian gauge dynamics on the cylinder maps exactly onto this integrable many-body system.

For completeness, we write the full Hamiltonian in the holonomy basis,
\begin{align}\label{eq:finalHop_eigenvalue_basis}
\begin{split}
    H&=-\frac{1}{2m}\left(\frac{d}{dz}-\frac{i}{r}\sum_{m=1}^{N-1}Q_0^m(y^m-y^{m+1})\right)^2-\frac{1}{16\pi^2m_{\rm YM}}\sum_{m=1}^N\left(\frac{d^2}{d(y^m)^2}-\sum_{n<m}
    \frac{d^2}{dy^mdy^n}\right)+V(\mathbf{y})\,,
\end{split}
\end{align}
with $V(\mathbf{y})$ as in \eqref{eq:CS-potential}. Note that there are crossed $y$-derivative terms. That is, even though the eigenvalue basis simplifies the potential term, it does not render the kinetic term diagonal.


\subsection{The space of gauge orbits}
\label{sec:gauge_orbits}

The fact that the gauge directions are compact introduces Gribov copies. Consider a gauge transformation that winds around the spatial circle
\begin{equation}
    U(t,x)=\exp(-i\Lambda^i(t,x)H_i)\,.
\end{equation}
Single-valuedness of $U$ then implies that $\Lambda^i$ needs to satisfy
\begin{equation}
    \Lambda^a(t,x=2\pi r) = \Lambda^a(t,x=0)+
    2\pi n^a\,, \quad n^a\in\mathbb{Z}\,.
\end{equation}
A simple choice is the linear ansatz
$\Lambda^a(t,x)=\dfrac{n^a x}{r}$. Under such a transformation, the spatial component of the gauge field $A_1=A_1^iH_i$ transforms as
\begin{equation}
    A_1 \to UA_1U^\dagger+ iU\partial_1U^\dagger= A_1 - 
    \frac{n^i}{r}H_i \,,
\end{equation}
where we used that $U$ and the Cartan generators commute. Thus the Cartan components $A_1^i$ are shifted by integer multiples of $1/r$.

Configurations that differ by such shifts are physically equivalent. They represent the same gauge orbit. Consequently, each $A_1^i$ can be restricted to a fundamental interval of length $1/r$, for instance  $0\leq A_1^i \leq 1/r$.
In terms of the dimensionless Cartan variables $Y^i$, see \eqref{eq:Acartan}, this can be equivalently written as $Y^i\sim Y^i+n^i$, $n^i \in \mathbb{Z}$ and we may choose the fundamental domain
\begin{equation}
    0\leq Y^i<1\,, \qquad i=1,\dots,N-1\,.
\end{equation}

Geometrically, this means that the classical configuration space of the Cartan component is compactified from $\mathbb{R}^{N-1}$ to the $(N-1)$-torus,
\begin{equation}\label{eq:RtoT}
    T_{SU(N)}=\frac{\mathbb{R}^{N-1}}{\mathbb{Z}^{N-1}}\,.
\end{equation}

In addition to these identifications, there is a further residual symmetry that must be accounted for: the Weyl group of $SU(N)$. The Weyl group acts by permuting the eigenvalues of the Wilson loop variable $\mathfrak{q}$; see \eqref{eq:qCartan}. It is isomorphic to the symmetric group $S_N$ and its action is detailed in Appendix~\ref{app:weyl}. Identifying configurations related by Weyl reflections yields the true space of gauge orbits,
\begin{equation}\label{eq:orbifold}
    \frac{T_{SU(N)}}{S_N}\,.
\end{equation}
The two quotients operations, the discrete translations $\mathbb{Z}^{N-1}$ that build the torus, and the permutations $S_N$, combine to form the semi-direct product $\mathbb{Z}^{N-1}\rtimes S_N$, because a Weyl reflection maps an integer shift to another integer shift: $\pi n^i\pi^{-1}\in \mathbb{Z}^{N-1}$ for all $n^i\in \mathbb{Z}^{N-1}$ and $\pi \in S_N$.

The equivalent configurations that arise from these identifications are known as Gribov copies or Gribov ambiguities. Because the space of gauge orbits
\eqref{eq:orbifold} has the structure of an orbifold \cite{Dixon:1986jc}, there is no way to completely eliminate these copies by a gauge-fixing procedure. 

\section{Quantization}
\label{sec:Quantization}

Our initial field theory was reduced to a mechanical system. Conjugate pairs are $(z,p_z)$, concerning the charged particle's degrees of freedom, and $(Y^i,P_0^j)$, related to the Wilson loop variable $\mathfrak{q}$. 

To quantize the system, note that, as a consequence of \eqref{eq:pqcommutator}, it follows that $P_0^j$ and $\mathfrak{q}$ satisfy
\begin{equation}\label{eq:Phatqcommutator}
    [P_0^j,\mathfrak{q}]=-\mathfrak{q} \frac{1}{2}(b^{-1})^{jk} H_k\,.
\end{equation}
This means that the momentum operator $P_0^j$ may be realized as
\begin{equation}\label{eq:P_operator}
    P_0^j = -\frac{i}{4\pi}\frac{\partial}{\partial Y_j}=-\frac{i}{4\pi}(b^{-1})^{jk}\frac{\partial}{\partial Y^k}\,.
\end{equation} 
We represent the momentum operator of the particle, $p_z$, as 
\begin{equation}
    p_z = -i\frac{d}{dz}\,.
\end{equation} 

The Hamiltonian operator obtained from \eqref{eq:finalH} is then
\begin{align}\label{eq:quantum_hamiltonian}
\begin{split}
    H&=-\frac{1}{2m}\left(\frac{d}{dz}-\frac{i}{r}b_{ij}Y^i Q_{0}^j\right)^2-\frac{(b^{-1})^{ij}}{16\pi^2m_{\rm YM}}\frac{\partial^2}{\partial Y^i\partial Y^j}+\frac{1}{2m_{\rm YM}}V(\mathbf{Y})\,,
\end{split}
\end{align}
where ${\bf Y}$ is now a set of operators acting on a Hilbert space.

Here, a preliminary analysis of the spectrum of the Hamiltonian is in order. Consider the limit of $r\to\infty$. In this limit pure Yang-Mills theory is known to become trivial \cite{Rajeev:1988zb}. In our setup this can be seen as follows. The Hamiltonian reduces to
\begin{equation}
    H\sim \frac{1}{m_{\rm YM}}P_0^i P_{0i}\,, \qquad \frac{1}{m_{\rm YM}}=g_{\rm YM}^2 \pi r\,.
\end{equation}
Thus, the energy eigenvalues are proportional to radius $r$ of the cylinder multiplied by the quadratic Casimir of the representation of the gauge group. Consequently, in this limit, all eigenenergies diverge except that of the ground state, which has zero Casimir and therefore vanishing energy. The theory possesses only one finite-energy state and becomes trivial. 

In contrast, when coupled to a charged particle, the $r\to\infty$ limit yields a Hamiltonian of the form
\begin{equation}\label{eq:Hatlarger}
    H\sim \frac{1}{m_{\rm YM}}\left(P_0^iP_{0i}+\frac{1}{2}V(\mathbf{Y})\right) \,.
\end{equation}
The potential $V(\mathbf{Y})$ survives the limit and thus modifies the spectrum. In this case, because the potential has singularities, one must ensure that the quantum Hamiltonian is self-adjoint, typically requiring a choice of boundary conditions~\cite{Asorey2013Singular}. Relevant discussions in related contexts can be found in~\cite{deAlfaro:1976vlx,Camblong:2000ec,Beane:2000wh,Kaplan:2009kr,daSilva:2023asj,daSilva:2025vkl,Case:1950an,Gupta:1993id}.


\subsection{General structure of eigenstates}
\label{subsec:general_structure_eigenstates}

The Hamiltonian \eqref{eq:quantum_hamiltonian} is defined on the manifold $z \in [0, 2\pi r)$ and $Y^i \in [0,1)$  for $i = 1,\dots, N-1$. For the eigenvalue problem $H\psi(z,\mathbf{Y})=E\psi(z,\mathbf{Y})$, the wavefunction $\psi(z,\mathbf{Y})$ satisfies periodic boundary conditions in $z$,
\begin{equation}\label{eq:boundary_condition_z}
\psi(z+2\pi r, \mathbf{Y}) = \psi(z, \mathbf{Y})\,,
\end{equation}
while in the $Y^i$ directions,
\begin{equation}\label{eq:boundary_condition_Y}
\psi(z, \mathbf{Y} + {\rm e}^i) = e^{i b_{ij}Q_{0}^jz/r} \, \psi(z, \mathbf{Y}), \qquad i=1,\dots,N-1,
\end{equation}
where ${\rm e}^i$ is a vector representing a unit length translation in the $i$-th direction. That is, $\mathbf{Y} + {\rm e}^i$ stands for $\{Y^1,\dots,Y^i+1,\dots,Y^{N-1}\}$.

\begin{figure}[h]
\centering
\begin{tikzpicture}[scale=1.2, every node/.style={font=\small}]

\begin{scope}[xshift=3.5cm]

  \node[right] at (3.2,-0.75) {$T_{Y^i}$};
  \node[below] at (1.5,-1.6) {$T_z$};
  \node[left] at (-0.2,-0.75) {$T_{Y^i}$};
  \node[above] at (1.5,0.1) {$T_z$};
  
  \draw[cyan, thick,->] (0,-1.5) -- (3,-1.5);
  \draw[cyan, thick,->] (3,-1.5) -- (3,0);
  \draw[orange, thick,->] (0,0) -- (3,0);
  \draw[orange, thick,->] (0,-1.5) -- (0,0);

  \draw[black] (1.55,-1.45) -- (1.45,-1.55);
  \draw[black] (1.45,-0.05) -- (1.55,0.05);
  \draw[black] (0.05,-0.7) -- (-0.05,-0.8);
  \draw[black] (0.05,-0.75) -- (-0.05,-0.85);
  \draw[black] (3.05,-0.7) -- (2.95,-0.8);
  \draw[black] (3.05,-0.75) -- (2.95,-0.85);
 
  \node[left] at (0,-1.5) {$(z,\mathbf{Y})$};
  \node[left] at (0,0) {$(z,\mathbf{Y}+{\rm e}^i)$};
  \node[right] at (3,0) {$(z+2\pi r,\mathbf{Y}+{\rm e}^i)$};
  \node[right] at (3,-1.5) {$(z+2\pi r,\mathbf{Y})$};
\end{scope}

\end{tikzpicture}
\caption{Two non-trivial cycles in configuration space: the horizontal cycle is the spatial $S^1$ factor from the cylindrical manifold and the vertical cycle is a large gauge transformation along $Y^i$ of $T_{SU(N)}$.}
\label{fig:2cycles}
\end{figure}
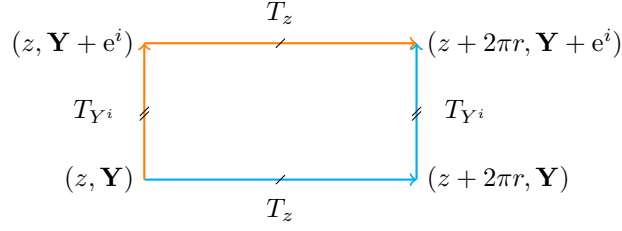

The boundary conditions lead to non-Abelian charge quantization. To see this, take a full cycle in the $z$ direction, generated by $T_z$, and pick a full cycle in one of the $\mathbf{Y}$ directions, generated by $T_{Y^i}$. There are two possible paths to follow to complete these cycles, as illustrated in Figure \ref{fig:2cycles}. 

If we complete a cycle in $z$ followed by a cycle in $Y^i$, we have
\begin{align}
\begin{split}
    T_{Y^i}T_z\psi(0,\mathbf{Y})&= T_{Y^i}\psi(2\pi r,\mathbf{Y})=e^{i2\pi b_{ij}Q_0^j}\psi(2\pi r,\mathbf{Y}+{\rm e}^i)\,.
\end{split}
\end{align}
Alternatively, if we complete the two cycles in the opposite order, we find
\begin{align}
\begin{split}
    T_zT_{Y^i}\psi(0,\mathbf{Y})&= T_{z}\psi(0,\mathbf{Y}+{\rm e}^i)=\psi(2\pi r,\mathbf{Y}+{\rm e}^i)\,.
\end{split}
\end{align}
Equivalence between the two paths implies non-Abelian charge quantization
\begin{equation}\label{eq:non-Abelian_charge_quantization}
    b_{ij}Q_0^j \,\in\mathbb{Z}\,, \quad i=1,\dots,N-1.
\end{equation}

Furthermore, using a Fourier expansion in $z$ we write
\begin{equation}\label{eq:psi}
\psi(z,\mathbf{Y}) = \sum_{n \in \mathbb{Z}} e^{i n z / r} \, \varphi_n(\mathbf{Y})\,.
\end{equation}
Substituting this into the eigenvalue equation yields a decoupled set of equations for each mode $n$:
\begin{equation}
\left( - \frac{(b^{-1})^{ij}}{16\pi^2 m_{\mathrm{YM}}} \frac{\partial^2}{\partial Y^i \partial Y^j}+ \frac{1}{2m r^2} \bigl( n - b_{ij}Y^i Q_0^j \bigr)^2  + \frac{1}{2m_{\mathrm{YM}}} V(\mathbf{Y}) \right) \varphi_n(\mathbf{Y}) = E \, \varphi_n(\mathbf{Y}).
\end{equation}
Note that the boundary condition \eqref{eq:boundary_condition_Y} implies that $\varphi$ satisfies
\begin{equation} \label{eq:bc_varphi} \varphi_n(\mathbf{Y}+{\rm e}^i)=e^{ib_{ij}Q_0^j z/r}\varphi_n(\mathbf{Y})\,.
\end{equation}
We can also obtain a spectral flow condition, that is, a shift in the momentum modes,
\begin{equation}\label{eq:spectral_flow}
\varphi_n(\mathbf{Y} + {\rm e}^i) =\varphi_{n-b_{ij}Q_{0}^j}(\mathbf{Y})\,.
\end{equation}

Note that, even though the Hamiltonian \eqref{eq:quantum_hamiltonian} is invariant under Weyl transformations $S_N$, the boundary condition equation \eqref{eq:boundary_condition_Y} is not. This signals the presence of an anomaly, as discussed in~\cite{Esteve:1986db,Balachandran:2011bv}.


\subsection{The case of $SU(2)$}
\label{subsec:su2}

The simplest non-Abelian group to be considered is the $SU(2)$. Despite its simplicity, it already displays interesting features.

First, we see that since there is only one Cartan generator and two root vectors  (see \eqref{eq:matricessu2}), there will be only one gauge field variable, one constant $Q_{0}^i$ and one constant $K^1$ appearing in the Hamiltonian. So, to write the Hamiltonian operator explicitly, we simply drop the indices: $Y^i\to Y$, $Q_0^i\to Q_0$, $K^1\to K$. Carefully using \eqref{eq:absu2}, we obtain
\begin{equation}\label{eq:Hsu2}
    H=\frac{1}{2m}\left(p_z-\frac{2}{r}Y Q_{0}\right)^2+\frac{1}{2m_{\rm YM}}\bigg(P_{0} P_{0}+\frac{K}{\sin^2\left(2\pi Y\right)}\bigg)\,.
\end{equation}

As discussed in Section~\ref{subsec:general_structure_eigenstates}, non-Abelian charge quantization implies $2Q_0\in\mathbb{Z}$. This implies that a $SU(2)$ charged particle may have integer or half-integer isospin. Furthermore, the eigenstates $\psi(z,Y)$ of the Hamiltonian take the form \eqref{eq:psi} which, in this case, leads to the following decoupled set of equations for each mode
\begin{equation}
\left(  - \frac{1}{32\pi^2 m_{\mathrm{YM}}} \frac{\partial^2}{\partial Y^2} +\frac{1}{2m r^2} \bigl( n - 2Y Q_0 \bigr)^2+ \frac{1}{2m_{\mathrm{YM}}} \frac{K}{\sin^2(2\pi Y)} \right) \varphi_n(Y) = E \, \varphi_n(Y).
\end{equation}
The $SU(2)$ realization of \eqref{eq:bc_varphi} leads to the boundary conditions
\begin{equation}
\varphi_n(Y+1)=e^{i2Q_0z/r}\varphi_n(Y)\label{eq:bc_varphi_su2}.
\end{equation}
The spectral flow condition \eqref{eq:spectral_flow} becomes 
\begin{equation}
    \varphi_n(Y+1)=\varphi_{n-2Q_0}(Y)\,. \label{eq:spectral_flow_su2}
\end{equation}

Finally, we comment on the potential $V(Y)=\frac{K}{\sin^2(2\pi Y)}$. It has a periodicity of $1/2$. Therefore, the fundamental unit cell is a 1-simplex, a segment of length $1/2$, as shown in Figure \ref{fig:potentialsu2} below.

\begin{figure}[H]
     \centering
     \includegraphics[width=0.45\textwidth]{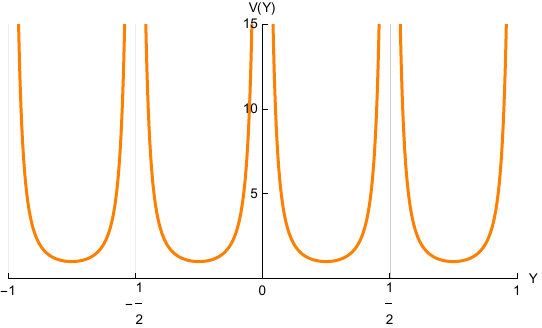}

    \caption{Potential $V(Y)$ for $G=SU(2)$ and $K=1$. Other values for $K$ simply shift the curves vertically.} 
    \label{fig:potentialsu2}
\end{figure}


\subsection{The case of $SU(3)$}

For $SU(3)$, there are two Cartan generators and six root vectors. Correspondingly, we have two gauge variables $Y^1, Y^2$, and two plus three constant charges: $Q_0^1, Q_0^2$ and $K^1, K^2, K^3$. Using the group-theoretic data listed in \eqref{eq:aandbSU(3)}, the Hamiltonian \eqref{eq:finalH} becomes explicit.

The kinetic and minimal coupling terms are
\begin{equation}
P_0^i P_{0i} = \left(-\frac{i}{4\pi}\right)^2 \frac{2}{3}\left( \frac{\partial^2}{\partial Y^1\partial Y^1} + \frac{\partial^2}{\partial Y^1\partial Y^2} + \frac{\partial^2}{\partial Y^2\partial Y^2} \right),
\label{eq:SU3_kinetic}
\end{equation}
and
\begin{equation}
\frac{1}{r} b_{ij} Y^i Q_0^j = \frac{1}{r}\Big( Q_0^1 (2Y^1 - Y^2) + Q_0^2 (-Y^1 + 2Y^2) \Big).
\label{eq:SU3_minimal}
\end{equation}
The potential is
\begin{equation}
V(Y^1, Y^2) = \frac{K^1}{\sin^2\big(\pi(2Y^1 - Y^2)\big)} + \frac{K^2}{\sin^2\big(\pi(Y^1 + Y^2)\big)} + \frac{K^3}{\sin^2\big(\pi(-Y^1 + 2Y^2)\big)}.
\label{eq:SU3_potential}
\end{equation}

The coordinates are initially restricted to
\begin{equation}
0 \le z < 2\pi r,\qquad 0 \le Y^1, Y^2 < 1,
\label{eq:SU3_domain}
\end{equation}
but the true gauge orbit space is the further quotient by the Weyl group $S_3$,
\begin{equation}
\frac{T_{SU(3)}}{S_3}.
\label{eq:SU3_orbifold}
\end{equation}

The Weyl group permutes the eigenvalues of the Wilson loop $\mathfrak{q}$. On the $Y^i$ coordinates, it is generated by $\mathcal{P}_1$ and $\mathcal{P}_2$ outlined in \eqref{eq:P_su3}. The Hamiltonian is invariant under this action (as demonstrated in Appendix~\ref{app:weyl}), so the physical domain is the 2-simplex (triangle) obtained from the $\mathbf{Y}$-square \eqref{eq:SU3_domain} by the $S_3$ quotient. 

\begin{figure}[h!]
    \centering
    \subfigure[]{\includegraphics[width=0.34\textwidth]{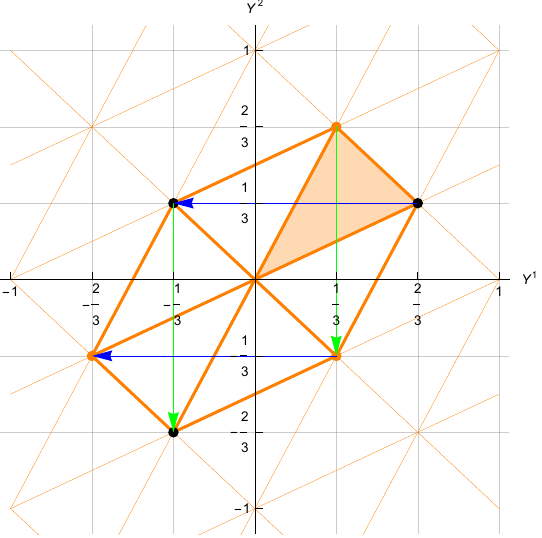}\label{subfig:su3lattice}} \hspace{2cm}
    \subfigure[]{\includegraphics[width=0.4\textwidth]{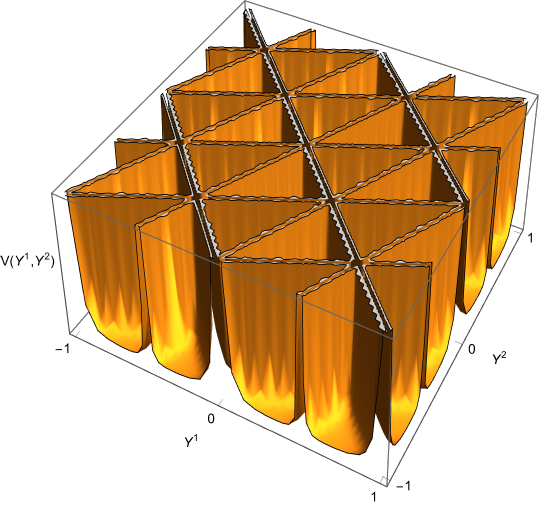}\label{subfig:su3potential}} 
    \caption{$G=SU(3)$: \subref{subfig:su3lattice} the 2D lattice from $T_{SU(3)}/S_3$. The fundamental unit cell is 2-simplex, a triangle. The honeycomb lattice is obtained after acting upon its vertices with permutation matrices. Blue and green shifts correspond to a $\mathcal{P}_1$ and a $\mathcal{P}_2$ action, respectively. The entire plane is covered by copies of this hexagon structure, generated by large gauge transformations. In \subref{subfig:su3potential} the potential $V(Y^1,Y^2)$ assuming $K^{\hat\imath}=1$, $\hat\imath=1,2,3$.}
    \label{fig:potentialSU3}
\end{figure}

The two-dimensional lattice from $T_{SU(3)}/S_3$ is shown in Figure~\ref{subfig:su3lattice}. The fundamental triangle (chosen to be the one filled in orange) and its Weyl images form a regular hexagon: $\mathcal{P}_2$ maps the upper orange vertex to the lower right and left orange vertices, and $\mathcal{P}_1$ maps the rightmost black vertex to the upper and lower left black vertices. The plane is then covered by translations of this hexagon generated by large gauge transformations (the $\mathbb{Z}^2$ lattice).

The potential is singular on the hyperplanes
\begin{equation}
(2Y^1 - Y^2)\,, (Y^1 + Y^2)\,, (-Y^1 + 2Y^2) \in \mathbb{Z}\,. 
\label{eq:SU3_singularities}
\end{equation}
These are the Weyl reflection planes. Their intersections with the torus form the boundaries of the Weyl chamber. Figure~\ref{subfig:su3potential} plots the potential $V(Y^1, Y^2)$ over several torus copies: the singular ridges follow the lattice lines, while the smooth triangular regions are the Weyl chambers.

\section{Discussion}
\label{sec:discussion}

In this work we studied Yang-Mills theory on a cylinder coupled to a point particle. We found that for $SU(N)$ gauge group, a proper treatment of the gauge degrees of freedom leads to a Hamiltonian description of a one-dimensional quantum many-body system. The interaction consists of a singular potential of the Calogero-Sutherland type~\cite{Calogero:1970nt,Sutherland:1971kq} plus an additional harmonic term originating from the minimal coupling between the particle's position and the gauge field. 

The singular Calogero-Sutherland potential gives rise to a crystal structure whose fundamental cell is an 
$(N-1)$-simplex. The configuration space is determined by the Weyl chamber, whose images under Weyl transformations and large gauge transformations generate a lattice. We illustrated these structures for $SU(2)$ and $SU(3)$. In the former, the unit cell is a segment of length $1/2$. In the latter, it is a triangle, and the action of the Weyl group produces a honeycomb lattice.

Our results lay the groundwork for analyzing the entanglement entropy between the particle and gauge degrees of freedom, naturally extending the Abelian treatment of~\cite{Ares:2019oil} to non-Abelian gauge groups. This development fits into a broader program of studying entanglement in the presence of gauge symmetries. Entanglement can probe the physical Hilbert space: it can characterize topological order~\cite{Kitaev:2005dm,Levin:2006zz,Chen:2010gda}, illuminate the emergence of locality in constrained systems~\cite{Balachandran:2013a, Balachandran:2013b}, and reveal algebraic and entropic aspects of gauge theories~\cite{Balachandran:2015a, Balachandran:2015b, Balachandran:2013c, Balachandran:2013d}. Moreover, entanglement plays a central role in the emerging space paradigm~\cite{VanRaamsdonk:2010pw,Maldacena:2013xja}, where spacetime geometry arises from patterns of quantum entanglement; in this context, large-volume entanglement in bipartite states of 2D Yang-Mills was recently investigated in~\cite{Melnikov:2026adx}.

In our setting, we observe that the Hamiltonian is invariant under the action of the Weyl group, whereas the boundary conditions on the wavefunctions are not. This signals the presence of an anomaly and calls for a proper identification of the anomaly-free physical state~\cite{Esteve:1986db,Balachandran:2011aa}. This is a crucial ingredient for a subsequent entanglement study.


\section*{Acknowledgments}

We thank Dmitry Melnikov for useful discussions. The work of M.T. was supported by the Simons Foundation through award number 1023171-RC and by the Brazilian National Council for Scientific and Technological Development (CNPq) through grant 445944/2024-2. M.T. also acknowledges IIF-FINEP grant 213/2024. The work by A.R.Q. was supported by CNPq under process number 310533/2022-8. A.R.Q. also acknowledges FAPESQ-PB.

\appendix

\section{Group theory conventions and notation}
\label{app:grouptheory}

We follow the conventions of \cite{Langmann:1992np}. The internal symmetry group $G$ is a compact connected Lie group. In particular, we consider $G=SU(N)$. The associated Lie algebra $\frak{g}$ is simple and generators $T^a$, $a=1,...,N^2-1$, satisfy
    \begin{equation}
        (T^a)^\dagger = T^a\,,\qquad \Tr(T^a T^b) = \frac{1}{2} \delta^{ab}\,,\qquad
        \left[T^a,T^b\right] = i f^{abc} T^c\,,
        \end{equation}
where $f^{abc}$ are the structure constants of the algebra. In terms of the generators, an element $\Lambda$ of the algebra $\mathfrak{g}$ is decomposed as $\Lambda=\Lambda^a T^a$, while an element $U$ of the group $G$ is obtained via exponentiation $U=\exp(i\Lambda)$.

Alternatively, elements of the simple Lie algebra $\mathfrak{g}$ may be written in the Cartan-Weyl basis containing $N-1$ Cartan generators, $\{H_i,i=1,\dots,N-1\}$, plus a complementary space of $N(N-1)$ root vectors, $\{E^\pm_{\hat\imath},{\hat\imath}=1,\dots,\frac{1}{2}N(N-1)\}$:
\begin{equation}
    \Lambda= \Lambda_0^{i}H_i +  \left( \Lambda_+^{\hat\imath}E_{\hat\imath}^- + \Lambda_-^{{\hat\imath}} E_{\hat\imath}^+ \right)\,.
\end{equation}
As in the main body of the paper, summation over repeated indices is assumed unless otherwise stated.

In terms of an $N\times N$ matrix $e_{ij}$ of elements $(e_{ij})_{kl}=\delta_{ik}\delta_{jl}$, the base elements of such a decomposition may be written as $H_i=e_{i,i}-e_{i+1,i+1}$, 
$E_1^+=e_{1,2}$, $E_2^+=e_{1,3}$, $\cdots$, $E_{N-1}^+=e_{1,N}$, $E_N^+=e_{2,3}$, $\cdots$, $E_{\frac{1}{2}N(N-1)}^+=e_{N-1,N}$ and $E_{\hat\imath}^-=(E^+_{\hat\imath})^T$. They satisfy
\begin{align}
        [H_i,H_j]&=0\,, \label{eq:PropertiesCartanWeyl1} \\
        [H_i,E_{{\hat\imath}}^\pm]&=\pm a_{i{\hat\imath}}E_{\hat\imath}^\pm \,,\label{eq:PropertiesCartanWeyl2}
\end{align}
where $a_{i{\hat\imath}}=\delta_{ik({\hat\imath})}-\delta_{il({\hat\imath})}-\delta_{i+1,k({\hat\imath})}+\delta_{i+1,l({\hat\imath})}$, with $k({\hat\imath}),l({\hat\imath})$ determined from $E_{\hat\imath}^+=e_{k({\hat\imath}),l({\hat\imath})}$. In~\eqref{eq:PropertiesCartanWeyl2}, there is no implicit sum over $\hat\imath$.
They also satisfy
\begin{align}
        {\rm Tr}(H_i H_j)&=b_{ij}\,,\label{eq:PropertiesCartanWeyl3} \\
        {\rm Tr}(E_{\hat\imath}^+ E_{\hat\jmath}^-)&=\delta_{{\hat\imath}{\hat\jmath}}\,, \label{eq:PropertiesCartanWeyl4}\\
        {\rm Tr}(H_i E_{\hat\imath}^\pm)&={\rm Tr}(E_{\hat\imath}^\pm E_{\hat\jmath}^\pm)= 0\,, \label{eq:PropertiesCartanWeyl5}
\end{align}
where $b_{ij}=2\delta_{ij}-\delta_{i+1,j}-\delta_{i,j+1}$. 

The matrix $a$ is the root matrix. Its elements $a_{i{\hat\imath}}$ are components of the roots in the basis dual to the Cartan generators $H_i$: $a_{i{\hat\imath}}=\alpha_{\hat\imath}(H_i)$, where ${\hat\imath}$ labels the root $\alpha_{\hat\imath}$ and $i$ the Cartan generator $H_i$. The set of vectors $\{\alpha_{\hat\imath}\}$ constitutes the root system of the algebra. The matrix of elements $b_{ij}$ is the Cartan matrix of the Lie algebra of $SU(N)$. In the Cartan classification, it encodes the inner products of the simple roots \cite{Georgi:1999wka}.

Elements of the Cartan-Weyl basis satisfy
\begin{align}\label{eq:expCartanWeylexp}
    \begin{split}
        \exp\left(-i \Lambda^i H_i\right)H_j \exp\left(i \Lambda^i H_i\right) &= H_j \,,\\
        \exp\left(-i \Lambda^i H_i\right)E_{\hat\imath}^\pm \exp\left(i \Lambda^i H_i\right) &= E_{\hat\imath}^\pm \exp\left(\mp i \Lambda^i a_{i{\hat\imath}}\right)\,,
    \end{split}
\end{align}
where the index $\hat\imath$ is not summed over.

For convenience, we write the generators, as well as $a_{i{\hat\imath}}$ and $b_{ij}$ elements explicitly for the $G=SU(2)$ and $G=SU(3)$ cases. In the former case, the Cartan-Weyl basis is expanded by a single Cartan generator $H$ and two root vectors $E^\pm$:
\begin{equation}\label{eq:matricessu2}
    \begin{split}
        H=\begin{pmatrix}
            1 & 0 \\
            0 & -1
        \end{pmatrix}\,, \quad E^+=\begin{pmatrix}
            0 & 1 \\
            0 & 0
         \end{pmatrix}\,, \quad E^-=\begin{pmatrix}
             0 & 0 \\
             1 & 0
         \end{pmatrix}\,.
    \end{split}
\end{equation}
There is one element $b_{11}$ and one element $a_{11}$, both equal to 2,
\begin{equation}\label{eq:absu2}
    b_{11}=2\,,\quad a_{11}=2\,.
\end{equation}
When $G=SU(3)$, the Cartan-Weyl basis is instead expanded by two Cartan generators $H_i$,
\begin{align}
        H_1 &= \begin{pmatrix}
            1 & 0 & 0 \\
            0 & -1 & 0 \\
            0 & 0 & 0
        \end{pmatrix}\,, \qquad H_2 = \begin{pmatrix}
            0 & 0 & 0 \\
            0 & 1 & 0 \\
            0 & 0 & -1
        \end{pmatrix}\,,
\end{align}
and six root vectors $E_{\hat\imath}^\pm$,
\begin{align}
    \begin{split}
         E_1^+ &= \begin{pmatrix}
            0 & 1 & 0 \\
            0 & 0 & 0 \\
            0 & 0 & 0
        \end{pmatrix}\,, \quad E_2^+ = \begin{pmatrix}
            0 & 0 & 1 \\
            0 & 0 & 0 \\
            0 & 0 & 0
        \end{pmatrix}\,, \quad E_3^+ = \begin{pmatrix}
            0 & 0 & 0 \\
            0 & 0 & 1 \\
            0 & 0 & 0
        \end{pmatrix}\,, \quad E_{\hat\imath}^-=(E_{\hat\imath}^+)^T\,, {\hat\imath}=1,2,3\,.
    \end{split}
\end{align}
The $2\times3$ matrix of elements $a_{i{\hat\imath}}$ and  the $2\times 2$ matrix of elements $b_{ij}$ are given by
\begin{equation}\label{eq:aandbSU(3)}
    a=\begin{pmatrix}
        2 && 1 && -1 \\
        -1 && 1 && 2
    \end{pmatrix}\,, \qquad 
    b=\begin{pmatrix}
        2 && -1 \\
        -1 && 2
    \end{pmatrix}\,.
\end{equation}


\section{Solving the Gauss law constraint}
\label{app:gauss}

In this appendix, we collect the computational details of the rotation to the Cartan basis mentioned in Section \ref{subsec:cartanbasis}.

The first object we analyze is the isotopic spin $I(t)$. First, note that \eqref{eq:DtauI} for $A_0=0$ gives
\begin{equation}\label{eq:DtIapp}
    \frac{dI(t)}{dt}-i\dot{z}[A_1,I(t)]=0\,.
\end{equation}
Decomposing $I(t)$ in the Cartan-Weyl basis, 
\begin{equation}
    I(t)=I_0^{i}(t)H_i+I_+^{{\hat\imath}}(t)E_{\hat\imath}^- +I_-^{{\hat\imath}}(t)E_{\hat\imath}^+\,,
\end{equation}
and using $A_1=\frac{Y^i(t)H_i}{r}$, the properties listed in (\ref{eq:PropertiesCartanWeyl1}-\ref{eq:PropertiesCartanWeyl5}) imply that the commutator appearing on the left hand-side of \eqref{eq:DtIapp} becomes
\begin{equation}
    [A_1,I(t)]= \frac{Y^i(t)}{r} a_{i{\hat\imath}}\left(-I_+^{{\hat\imath}}(t)E_{\hat\imath}^- + I_-^{{\hat\imath}}(t)E_{\hat\imath}^+\right)\,.
\end{equation}
Then the solutions to \eqref{eq:DtIapp} are
\begin{align}\label{eq:Isolapp}
    \begin{split}
        \frac{dI_0^{j}(t)}{dt}&=0 \Longrightarrow I_0^{j}(t)= Q^{j}_0\,,
        \\
        \frac{dI_{\pm}^{\hat\imath}(t)}{dt}\pm i  \dot{z}\frac{Y^i(t)}{r}a_{i{\hat\imath}}I_{\pm}^{{\hat\imath}}(t)&=0 \Longrightarrow I_\pm^{\hat\imath}(t)=Q_\pm^{{\hat\imath}} \exp\left[{\mp \frac{i}{r} \int_0^t  a_{i{\hat\imath}}Y^i(t)\dot z\,dt}\right]\,.
    \end{split}
\end{align}
with no implicit summation over $\hat\imath$.

In the same spirit, we consider the effect of the rotation to the Cartan basis upon $R(x)$. Recall that
\begin{equation}
    R(x) = \int_0^x dy\, \tilde{\rho}(y) = \int_0^x dy\, S(y)^{-1}\rho(y)S(y)\,.
\end{equation}
Expanding $\rho(x)$ in Fourier modes, with
\begin{equation}
    \varrho(n) =\int_0^{2\pi r} dx \,\rho(x) e^{-inx/R} \quad \text{and}\quad \varrho(n)= \varrho_0^{j}(n) H_j + \varrho_+^{{\hat\imath}}(n) E_{\hat\imath}^- + \varrho_-^{{\hat\imath}}(n) E_{\hat\imath}^+\,,
\end{equation}
we then make use of \eqref{eq:expCartanWeylexp} to write $R(x)$ as follows
\begin{align}
    \begin{split}
        R(x) &= \int_0^x dy\, e^{i\, Y^iH_i y/r}\left(\frac{1}{2\pi r}\sum_{n\in\mathbb{Z}} \varrho(n)e^{iny/r}\right)e^{-i\,Y^i H_i y/r}\\
        &=\int_0^{x} dy\, \sum_{n} \left( \varrho_0^{i}(n)H_i + \varrho_+^{{\hat\imath}}(n) E_{\hat\imath}^- e^{-i\, Y^ia_{i{\hat\imath}} y/r} + \varrho_-^{{\hat\imath}} E_{\hat\imath}^+ e^{i \, Y^i a_{i{\hat\imath}} y/r} \right)  \frac{e^{iny/r}}{2\pi r}\\
        &= \varrho_0^{i}(0)H_i\, \frac{x}{2\pi r} + \sum_{n\neq 0} \varrho_0^i(n)H_i \frac{i(1-e^{-inx/r})}{2\pi n}  \\
        &\hspace{0.5cm}+ \sum_n\bigg( \varrho_+^{{\hat\imath}}(n)E_{\hat\imath}^- \frac{i(1-e^{i(n- Y^ia_{i{\hat\imath}})x/r})}{2\pi(n-Y^ia_{i{\hat\imath}})}+\varrho_-^{{\hat\imath}}(n)E_{\hat\imath}^+ \frac{i(1-e^{i(n+ Y^ia_{i{\hat\imath}})x/r})}{2\pi(n+Y^ia_{i{\hat\imath}})}\bigg) \bigg]\,.
    \end{split}
\end{align}

Analogously, $\bar{R}(x)$ may be written as
\begin{align}
    \begin{split}
        \bar{R}(x)&= R(x)-\frac{1}{2\pi r} \int_0^{2\pi R}dy\,R(y)\\
        &= \varrho_0^{i}(0) H_i\frac{(x-\pi r)}{2\pi r} + \sum_{n\neq 0} \varrho_0^{i}(n)H_i\frac{(-ie^{inx/r})}{2\pi n}\\
        &\hspace{0.5cm}+ \sum_n \varrho_+^{{\hat\imath}}(n)E_{\hat\imath}^- \frac{i}{2\pi(n-Y^ia_{i{\hat\imath}})}\bigg(-e^{i(n- Y^ia_{i{\hat\imath}})x/r}-\frac{i}{2\pi}\frac{(e^{-2\pi i Y^ia_{i{\hat\imath}}}-1)}{n- Y^ia_{i{\hat\imath}}}\bigg)\\
        &\hspace{0.5cm}+  \sum_n \varrho_-^{{\hat\imath}}(n)E_{\hat\imath}^+ \frac{i}{2\pi(n+ Y^ia_{i{\hat\imath}})}\bigg(-e^{i(n+ Y^ia_{i{\hat\imath}})x/r}-\frac{i}{2\pi}\frac{(e^{2\pi i Y^ia_{i{\hat\imath}}}-1)}{n+ Y^ia_{i{\hat\imath}}}\bigg)\,.
    \end{split}
\end{align}
Thus, its components are
\begin{align}\label{eq:Rbarcompapp}
    \begin{split}
        \bar{R}_0^i(x)&= \rho_0^{i}(0) \frac{(x-\pi r)}{2\pi r} + \sum_{n\neq 0} \varrho_0^{i}(n)\frac{(-ie^{inx/r})}{2\pi n}\,,\\
        \bar{R}_+^{\hat\imath}(x)&= \sum_n \varrho_+^{{\hat\imath}}(n) \frac{i}{2\pi(n- Y^ia_{i{\hat\imath}})}\bigg(-e^{i(n-Y^ia_{i{\hat\imath}})x/r}-\frac{i}{2\pi}\frac{(e^{-2\pi i\, Y^ia_{i{\hat\imath}}}-1)}{n- Y^ia_{i{\hat\imath}}}\bigg)\,,\\
        \bar{R}_-^{\hat\imath}(x)&=  \sum_n \varrho_-^{{\hat\imath}}(n) \frac{i}{2\pi(n+ Y^ia_{i{\hat\imath}})}\bigg(-e^{i(n+Y^ia_{i{\hat\imath}})x/r}-\frac{i}{2\pi}\frac{(e^{2\pi i\, Y^ia_{i{\hat\imath}}}-1)}{n+ Y^ia_{i{\hat\imath}}}\bigg)\,.
    \end{split}
\end{align}

Plugging \eqref{eq:qCartan} and \eqref{eq:Rbarcompapp} into the constraint \eqref{eq:equivarianceconstraint}, we
obtain three independent equations,
\begin{align}\label{eq:HEmEpapp}
    \begin{split}
        H_i:&\quad \bar R_0^i(2\pi r) = \bar R_0^i(0)\,,\\
        E_{\hat\imath}^-:&\quad P_+^{\hat\imath}+ \bar R_+^{\hat\imath}(2\pi r) = \left( P_+^{\hat\imath} + \bar R_+^{\hat\imath}(0)\right)e^{-2\pi i \, Y^i a_{i{\hat\imath}}}\,,\\
        E_{\hat\imath}^+:&\quad  P_-^{\hat\imath} + \bar R_-^{\hat\imath}(2\pi r) = \left( P_-^{\hat\imath}+ \bar R_-^{\hat\imath}(0)\right)e^{2\pi i \,Y^i a_{i{\hat\imath}}}\,.
    \end{split}
\end{align}
Confronting these with \eqref{eq:Rbarcompapp}, we see that $\varrho_0^{i}(0)=0$ and therefore
\begin{equation}
    \bar R_0^i(x)=-i\sum_{n} \varrho_0^{i}(n)\frac{e^{inx/r}}{2\pi n}(1-\delta_{n,0})\,.
\end{equation}
Also, $P_\pm^{\hat\imath}$ becomes fully determined by $\bar{R}$, that is, by the charge distribution, via
\begin{align}\label{eq:Ppmapp}
    P_\pm^{\hat\imath}=\frac{1}{1-e^{\mp 2\pi i\, Y^ia_{i{\hat\imath}}}}\left(\bar{R}_\pm^{\hat\imath}(0)e^{\mp2\pi i \, Y^ia_{i{\hat\imath}}}-\bar{R}_\pm^{\hat\imath}(2\pi r) \right)\,.
\end{align}

Using \eqref{eq:Ppmapp} we obtain simple expressions for the combinations $P_\pm^{\hat\imath}+\bar{R}_\pm^{\hat\imath}(x)$, which appear in the Hamiltonian through $\tilde{\Pi}(x)=P+\bar{R}(x)$. We find
\begin{align}
    \begin{split}
        P_\pm^{\hat\imath} + \bar{R}_\pm^{\hat\imath}(x) &= -i \sum_n \varrho_\pm^{\hat\imath}(n)\frac{ e^{i(n\mp Y^ia_{i{\hat\imath}})x/r}}{2\pi(n\mp Y^ia_ {i{\hat\imath}})}\,.
    \end{split}
\end{align}

Now we can finally compute the electric field contribution to the Hamiltonian \eqref{eq:Htilded}. We have
\begin{align}\label{eq:trEtildeapp}
    \begin{split}
        \int_0^{2\pi r} dx\, {\rm Tr}(\tilde \Pi^2) &= \int_0^{2\pi r} dx\, {\rm Tr}\bigg\{\bigg[\left(P_0^i+ \bar{R}_0^i(x)\right)H_i + \left(P_+^{\hat\imath}+  \bar R_+^{\hat\imath}(x)\right)E_{\hat\imath}^- +\left( P_-^{\hat\imath} + \bar R_-^{\hat\imath}(x)\right) E_{\hat\imath}^+ \bigg]^2\bigg\}\\
        &= \int_0^{2\pi r} dx\,\bigg\{ \big(P_0^i+\bar{R}_0^i(x)\big)b_{ik}\big(P_0^k+\bar{R}_0^k(x)\big) + \big(P_+^{\hat\imath}+\bar{R}_+^{\hat\imath}(x)\big(P_-^{\hat\imath}+\bar{R}_-^{\hat\imath}(x)\big) \big)\\
        &\hspace{7.15cm}+ \big(P_-^{\hat\imath}+\bar{R}_-^{\hat\imath}(x)\big)\big(P_+^{\hat\imath}+\bar{R}_+^{\hat\imath}(x)\big)\bigg\}\\
        &= 2\pi r P_0^ib_{ik}P_0^k + \frac{r}{2\pi} \sum_{n} \bigg[ (1-\delta_{n,0})\frac{\varrho_0^{i}(n)b_{ik}\varrho^{0,k}(-n)}{n^2}+\frac{\varrho_+^{{\hat\imath}}(n)\varrho_-^{{\hat\imath}}(-n)}{(n-Y^ia_{i{\hat\imath}})^2} + \frac{\varrho_-^{{\hat\imath}}(n)\varrho_+^{{\hat\imath}}(-n)}{(n+Y^ia_{i{\hat\imath}})^2}\bigg]
    \end{split}
\end{align}
This result holds for any charge density $\rho(x)$. For the particular case of a charged point particle, $\rho(t,x)= I \delta(x-z)$, and we have
\begin{equation}
    \varrho(n)= \left(I_0^i(t)H_i + I_+^{\hat\imath}(t)E_{\hat\imath}^- + I_-^{\hat\imath}(t)E_{\hat\imath}^+\right) e^{-inz/r}\,.
\end{equation}
So \eqref{eq:trEtildeapp} becomes
\begin{align}\label{eq:trEtilde2app}
    \begin{split}
        \int_0^{2\pi r}dx\, {\rm Tr}(\tilde{\Pi}^2(x)) &= 2\pi r\,P_0^ib_{ik}P_0^k + \frac{r}{2\pi} \sum_{n} \Bigg[\frac{I_0^i(t)b_{ik}I_0^k(t)}{n^2}(1-\delta_{n,0})+\frac{I_+^{\hat\imath}(t)I_-^{\hat\imath}(t)}{(n- Y^ia_{i{\hat\imath}})^2} +\frac{I_-^{\hat\imath}(t)I_+^{\hat\imath}(t)}{(n+ Y^ia_{i{\hat\imath}})^2}\Bigg]\\
        &=2\pi r P_0^ib_{ik}P_0^k + \frac{\pi r}{2} \bigg( \frac{1}{3} I_0^i(t)b_{ik}I_0^k(t)  +\frac{I_+^{\hat\imath}(t)I_-^{\hat\imath}(t)+I_-^{\hat\imath}(t)I_+^{\hat\imath}(t)}{\sin^2(\pi Y^i a_{i{\hat\imath}})}\bigg)
    \end{split}
\end{align}
where we have used
\begin{align}
\begin{split}
    &\sum_{n=-\infty}^\infty \frac{(1-\delta_{n,0})}{n^2}=\frac{\pi^2}{3}\,,\\
    &\sum_{n=-\infty}^\infty \frac{1}{(n- Y^i a_{i{\hat\imath}})^2} =\sum_{n=-\infty}^{\infty} \frac{1}{(n+ Y^i a_{i{\hat\imath}})^2}=\frac{\pi^2}{\sin^2(\pi  Y^i a_{i{\hat\imath}})}\,.
\end{split}
\end{align}

Recalling that the components $I_0^i(t)$ and $I_\pm^{\hat\imath}$ satisfy \eqref{eq:Isolapp}, we see that not only $I_0^j$, but also the product $I_+^{\hat\imath}I_-^{\hat\imath}$ that appears in \eqref{eq:trEtilde2app} is time-independent. The electric field contribution yields
\begin{align}\label{eq:trEtilde3app}
    \begin{split}
        \int_0^{2\pi r}dx\, {\rm Tr}(\tilde{\Pi}^2)
        &= 2\pi r P_0^ib_{ik}P_0^k + \frac{\pi r}{2} \bigg(\frac{1}{3} Q_0^{i}b_{ik}Q_0^{k} + \frac{Q_+^{{\hat\imath}}Q_-^{{\hat\imath}}+Q_-^{{\hat\imath}}Q_+^{{\hat\imath}}}{\sin^2(\pi Y^i a_{i{\hat\imath}})}\bigg)\,,
    \end{split}
\end{align}
where $Q_0^{i}$ and $Q_\pm^{{\hat\imath}}$ are constants. Moreover, the term ${\rm Tr}(A_1(z) I)$ in the Hamiltonian  becomes
\begin{equation}\label{eq:TrAIapp}
    {\rm Tr}(A_1(z) I)=\frac{1}{r}{\rm Tr}\left[\left(Y^i H_ i\right)\left(I_0^jH_j+I_+^{\hat\imath}E_{\hat\imath}^-+I_-^{\hat\imath}E_{\hat\imath}^+\right)\right]
    =\frac{1}{r}Y^ib_{ij}Q_0^j=\frac{1}{r}Y^iQ_{0i}\,.
\end{equation}
Putting \eqref{eq:trEtilde3app} and \eqref{eq:TrAIapp} together, we arrive at the Hamiltonian presented in \eqref{eq:finalH}:
\begin{equation}
    H=\frac{1}{2m}\left(p_z-\frac{1}{r}Y^i Q_{0i}\right)^2+\frac{1}{m_{\rm YM}}P_{0}^i P_{0i}+\frac{1}{2m_{\rm YM}}\left(\frac{K^{\hat\imath}}{\sin^2(\pi Y^ia_{i{\hat\imath}})}\right)\,.
\end{equation}


\section{Weyl group action on the $SU(N)$ Hamiltonian}
\label{app:weyl}

The action of the Weyl group of $SU(N)$ is generated by elementary permutations $\pi_i$, $i=1,\dots,N-1$. Denote the eigenvalues of the Wilson loop variable $\mathfrak{q}$ as $(e^{2\pi iy^1},e^{2\pi iy^2},...,e^{2\pi iy^N})$. A permutation $\pi$ acts on the $y$-coordinates as
\begin{equation}
    y^i \to y^{\pi^{-1}(i)}\,.
\end{equation}
In terms of $Y^i$ variables, which are defined as cumulative sums of the $y^i$ according to \eqref{eq:Y_as_sum_of_y},
this action is implemented via matrices $\mathcal{P}$,
\begin{equation}\label{eq:permutation_P}
    Y^i \to Y^{i\ '}=\sum_j \mathcal{P}_j^iY^j\,, \quad \mathcal{P}_j^i=\sum_k \alpha_{\pi(k)}^i\beta_j^k\,.
\end{equation}
In this expression, $\alpha$ and $\beta$ are matrices read from the linear maps $Y^i=\sum_{j=1}^{N}\alpha^i_j y^j$ and $y^i=\sum_{j=1}^{N-1}\beta^i_j Y^j$. Their elements are:
\begin{equation}
    \alpha^i_j=\begin{cases}
        1,\quad j\leq i\\
        0,\quad \mbox{otherwise}
    \end{cases}\,, \qquad\qquad \beta_j^i=\begin{cases}
        \delta_j^1, \qquad\qquad i=1\\
        \delta_j^i-\delta_{j+1}^i,\quad 1<i<N\\
        -\delta_{j+1}^N, \qquad\,\, i=N
    \end{cases}\,.
\end{equation}

For the $K^{\hat\imath}$ variables, which live in the $N(N-1)/2$-dimensional space spanned by ordered vectors $e_{mn}, 1\leq m<n\leq N$, the Weyl transformation is given by
\begin{equation}\label{eq:permutation_K}
    K^{\hat\imath}\to K^{\hat\imath}\,'=\sum_{\hat\jmath}\hat{\mathcal{P}}^{\hat\imath}_{\hat\jmath}K^{\hat\jmath} \,,
\end{equation}
where $\hat{\mathcal{P}}$ acts on the basis vectors as $e_{mn}\to e_{\pi(m)\pi(n)}$. Concretely,
\begin{equation}
    \hat{\mathcal{P}}^{\hat\imath}_{\hat\jmath}= \begin{pmatrix}
        e_{\pi(1)\pi(1)}, &e_{\pi(1)\pi(2)}, & \dots & e_{\pi(1)\pi(N)}, & e_{\pi(2)\pi(3)}, & \dots & e_{\pi(2)\pi(N)}, & \dots &
 e_{\pi(N-1),\pi(N)}    \end{pmatrix}\,,
\end{equation}
with columns indexed by $\hat\jmath$ and rows by $\hat\imath$ in the chosen ordering of the $e_{mn}$.

\subsection{Invariance of the Hamiltonian}

The invariance of the harmonic oscillator term is shown as follows. Consider $Y^i Q_{0i}=Y^i b_{ij}Q_0^j$, where $Q_0^i$ transforms in the same way as $Y^i$. We have
\begin{equation}
    Y^i b_{ij}Q_{0}^j \to (\mathcal{P}^i_kY^k)b_{ij}(\mathcal{P}^j_lQ_0^l)=(\mathcal{P}^i_k b_{ij}\mathcal{P}_l^j)Y^kQ_0^l= b_{kl}Y^kQ_0^l\,.
\end{equation}
In the last equality, we used that $b$ is invariant under Weyl-reflection ($\mathcal{P}^T b \mathcal{P}=b$) which follows from $b$ being the Cartan matrix of $SU(N)$. Hence, the harmonic oscillator term is Weyl-invariant.

The same reasoning applies to $P_0^iP_{0i}$, which transforms into itself under the Weyl group. Thus the kinetic term is also invariant.

The potential term is
\begin{equation}\label{eq:VofY}
    V(\mathbf{Y})\equiv \sum_{{\hat\imath}=1}^{N(N-1)/2} \frac{K^{\hat\imath}}{\sin^2\left(\pi \displaystyle\sum_{i=1}^{N-1}Y^ia_{i{\hat\imath}}\right)}\,,\qquad {\bf Y}\equiv\{Y^i,\, i=1,\dots,N-1\}\,.
\end{equation}
where the coefficients $a_{i\hat\imath}$ encode the simple-root expansion of the positive roots. Under a Weyl transformation,
\begin{equation}
    V(\mathbf{Y})\to \sum_{\hat\imath=1}^{N(N-1)/2} \frac{\sum_{\hat\jmath}\hat{\mathcal{P}}^{\hat\imath}_{\hat\jmath}K^{\hat\jmath}}{\sin^2\left(\pi \sum_{i,j} \mathcal{P}^i_jY^j a_{i\hat\imath}\right)}\,.
\end{equation}
Individual terms are not invariant, but the sum over all $\hat\imath$ elements is invariant due to the reordering of positive roots induced by the permutation $\pi$. Therefore, $V(\mathbf{Y})$ is Weyl-invariant. 

\subsubsection*{Examples}

The analysis of Weyl invariance for $SU(2)$ is straightforward. The Weyl group is $\mathbb{Z}^2$ and acts by the simultaneous reflection $Y\to -Y$ and $Q_0\to -Q_0$, while the potential strength $K$ (there is only one root) remains unchanged. Under this transformation the kinetic term ($\propto P_Y^2$) and the potential term ($\propto K/\sin^2(2\pi Y)$) are manifestly invariant. The harmonic oscillator coupling ($\propto YQ_0$) also remains unchanged because the sign flip in $Y$ is compensated by the flip in $Q_0$. Thus the full Hamiltonian is invariant. 

We therefore turn to the more interesting case of $SU(3)$. The Weyl group is $S_3$, generated by two permutations $\pi_1,\pi_2$. In the $N-1=2$-dimensional space of $Y^i$, these act via
\begin{align}\label{eq:P_su3}
    \begin{split}
        \mathcal{P}_1=\begin{pmatrix}
            -1 & 1 \\ 0 & 1
        \end{pmatrix}\,,\qquad \mathcal{P}_2=\begin{pmatrix}
            1 & 0 \\ 1 & -1
        \end{pmatrix}\,.
    \end{split}
\end{align}
In the $N(N-1)/2=3$-dimensional space of $K^{\hat\imath}$, the corresponding matrices are
\begin{equation}
    \hat{\mathcal{P}}_1=\begin{pmatrix}
            1 & 0 & 0 \\ 0 & 0 & 1 \\
            0 & 1 & 0
        \end{pmatrix}\,,\qquad \hat{\mathcal{P}}_2=\begin{pmatrix}
           0 & 1 & 0 \\ 1 & 0 & 0 \\
           0 & 0 & 1
        \end{pmatrix}\,.
\end{equation}
The resulting transformations are:
\begin{align}
\pi_1:\begin{cases}
    2Y^1-Y^2\to -(2Y^1-Y^2)\\
    Y^1+Y^2 \leftrightarrow -Y^1+2Y^2\\[3pt]
    2Q_0^1-Q_0^2\to -(2Q_0^1-Q_0^2)\\
    Q_0^1+Q_0^2 \leftrightarrow -Q_0^1+2Q_0^2\\[3pt]
    K_2 \leftrightarrow K_3 
\end{cases}
\qquad
\pi_2:\begin{cases}
    2Y^1-Y^2\to Y^1+Y^2\\
    Y^1+Y^2\to 2Y^1-Y^2\\
    -Y^1+2Y^2\to Y^1-2Y^2\\[3pt]
    2Q_0^1-Q_0^2\to Q_0^1+Q_0^2\\
    Q_0^1+Q_0^2\to 2Q_0^1-Q_0^2\\
    -Q_0^1+2Q_0^2\to Q_0^1-2Q_0^2\\[3pt]
    K_1 \leftrightarrow K_2  
\end{cases} 
\end{align}
One can explicitly verify that these transformations leave each term of the $SU(3)$ Hamiltonian (kinetic, harmonic oscillator and potential) invariant. Hence the full Hamiltonian is Weyl-invariant, as claimed.

\clearpage
\bibliographystyle{JHEP}
\bibliography{refs}
\newpage 

\end{document}